\documentclass{article}
\pdfoutput=1
\usepackage[utf8]{inputenc}
\usepackage[numbers]{natbib}
\usepackage{graphicx} 
\usepackage{url}
\usepackage{hyperref}
\usepackage{authblk}
\usepackage{ulem}

\usepackage[margin=2cm]{geometry}

\title{Real-time 2019 Portuguese Parliament Election Results Dataset}

\author{Nuno Moniz\thanks{Email: nmmoniz@inesctec.pt}}
\affil{LIAAD - INESC Tec\\ FCUP - DCC, University of Porto}

\begin{document}

\maketitle

\abstract{This paper presents a data set describing the evolution of results in the Portuguese Parliamentary Elections of October 6$^{th}$ 2019. The data spans a time interval of 4 hours and 25 minutes, in intervals of 5 minutes, concerning the results of the 27 parties involved in the electoral event. The data set is tailored for predictive modelling tasks, mostly focused on numerical forecasting tasks. Regardless, it allows for other tasks such as ordinal regression or learn-to-rank.}

\vspace{0.4cm}

This dataset is available in the UCI Machine Learning Repository, with the name \textbf{Real-time Election Results: Portugal 2019}.


\section{Summary}\label{sec:summary}

The data set presented in this paper comprises the results of the Portuguese 2019 Parliamentary Elections for 4 hours and 25 minutes since the start of results coming public until the last parish was accounted. The electoral process had the participation of 27 parties, over 21 areas (global results plus 20 area results). Overall, the data set contains 21643 records over 28 features (including the target variable).

\subsection{Motivation}

Portugal has made results available in an online and updated fashion for many elections until today. However, from the perspective of predictive modelling, the data available raises several issues. The motivation for developing this data set relies on the fact that there is no public information concerning the order of such results - these are not published at the same time. Such detail adds a new level of information concerning the precedence of results and forecasting abilities depending on such level of information. The challenges and possibilities in such a level of information are the primary motivation of the data acquisition and curation process that led to this data set. The envisioned task for this process is numerical forecasting, attempting to assert the ability of significant anticipation of final results with a very high degree of confidence.


\section{Data Acquisition}\label{sec:dataacquisition}

The \textit{Secretaria-Geral do Ministério da Administração Interna} (SGMAI) handles the results of the electoral processes in Portugal. Such results are published online continuously, as they become available. The overall process on the communication/publication of results is as follows:

\begin{enumerate}
    \item Votes are counted when polls close;
    \item Once polling sections finish their counting, results are gathered and communicated to the SGMAI;
    \item Results are verified and published.
\end{enumerate}

Results of the 2019 Parliamentary Elections in Portugal are presented in the following website: \url{https://www.legislativas2019.mai.gov.pt/}. Additionally, we should stress that this website published what is considered provisional results. Final results are made available after a due process by the competent authorities. As such, this data set relates to such provisional results.

\subsection{Procedure}\label{subsec:methodology}

Information published on the official website by the SGMAI describe a nation-, district-, county- and parish-wide breakdown of results. Through the analysis of content and environment variables, it is possible to understand that results are stored in JSON files with a fixed-structure path as such: 

\begin{quote}
    \url{https://www.legislativas2019.mai.gov.pt/frontend/data/TerritoryResults?territoryKey=LOCAL-XXXXXX&electionId=AR&ts=}
\end{quote}

The respective ids of each location must replace the local id (\textbf{XXXXXX}), which is also made available in the content and environment variables of the website. 

For conciseness, we will not describe the schema of the information. Regardless, given its availability, it is relatively easy to grasp, and the accompanying code of this paper describes the information acquired.

We should note that there were two processes of data acquisition: an online and an offline process. In other words, some information (district-level information) is acquired while the results are published, and other information (parish-level information) acquired after the process. Due to the limitation of queries, the acquisition of parish-level information (3092 queries) would be too much of a burden to the results website.

\subsubsection{Online Acquisition}\label{subsubsec:onlineacquisition}

For the online process of data acquisition, while the results are published, the following methodology was applied.

\begin{enumerate}
    \item the identifiers of each voting district were collected (20), also, the global (national) results (total 21);
    \item results start to appear a few minutes after 8 PM (local time) on the 6$^th$ of October 2019, at which point, an automatic procedure to acquire the data was activated in 5-minute intervals;
    \item this process includes obtaining the JSON file containing the electoral results of each district (21);
    \item for each district, information is obtained concerning the statistics of the overall voting procedure, along with the individual level of voting for each party;
    \item new information is used to update the files containing the acquired data.
\end{enumerate}

The automatic procedure to acquire data was active from 6:56 PM (local time) of the 6$^th$ of October 2019, and 00:35 AM of the 7$^th$ of October 2019.

The application of this methodology resulted in two separate files: \textit{i)} overall statistics on the voting procedure and \textit{ii)} voting results for each party and each district. The description of each attribute of these files follows in Tables~\ref{tbl:overallresults} and \ref{tbl:votes}

\begin{table}[h]
\begin{center}
\scriptsize
\caption{Attributes collected in real-time concerning the overall district-level information of the electoral process.}\label{tbl:overallresults}
\begin{tabular}{l l p{8cm}}
\textbf{Variable}                & \textbf{Type}    & \textbf{Description} \\
\hline
\textbf{time}            & \textit{timestamp}  & Date and time of the data acquisition \\
\textbf{territoryFullName}            & \textit{string}  & Complete name of the location (district or nation-wide) \\
\textbf{territoryName}            & \textit{string}  & Short name of the location (district or nation-wide) \\
\textbf{territoryKey}            & \textit{string}  & Official identifying key of the location (e.g. \textit{LOCAL-500000}) \\
\textbf{totalMandates}            & \textit{numeric}  & MP's elected at the moment\\
\textbf{availableMandates}            & \textit{numeric}  & MP's left to elect at the moment\\
\textbf{numParishes}            & \textit{numeric}  & Total number of parishes in this location\\
\textbf{numParishesApproved}            & \textit{numeric}  & Number of parishes approved in this location\\
\textbf{blankVotes}            & \textit{numeric}  & Number of blank votes\\
\textbf{blankVotesPercentage}            & \textit{numeric}  & Percentage of blank votes\\
\textbf{nullVotes}            & \textit{numeric}  & Number of null votes\\
\textbf{nullVotesPercentage}            & \textit{numeric}  & Percentage of null votes\\
\textbf{votersPercentage}            & \textit{numeric}  & Percentage of voters\\
\textbf{subscribedVoters}            & \textit{numeric}  & Number of subscribed voters in the location\\
\textbf{totalVoters}            & \textit{numeric}  & Number of votes cast\\
\textbf{pre.totalMandates}            & \textit{numeric}  & MP's elected at the moment (previous election)\\
\textbf{pre.availableMandates}            & \textit{numeric}  & MP's left to elect at the moment (previous election)\\
\textbf{pre.blankVotes}            & \textit{numeric}  & Number of blank votes (previous election)\\
\textbf{pre.blankVotesPercentage}            & \textit{numeric}  & Percentage of blank votes (previous election)\\
\textbf{pre.nullVotes}            & \textit{numeric}  & Number of null votes (previous election)\\
\textbf{pre.nullVotesPercentage}            & \textit{numeric}  & Percentage of null votes (previous election)\\
\textbf{pre.votersPercentage}            & \textit{numeric}  & Percentage of voters (previous election)\\
\textbf{pre.subscribedVoters}            & \textit{numeric}  & Number of subscribed voters in the location (previous election)\\
\textbf{pre.totalVotes}            & \textit{numeric}  & Percentage of blank votes (previous election)\\
\hline
\end{tabular}
\end{center}
\end{table}

\begin{table}[h]
\begin{center}
\scriptsize
\caption{Attributes collected in real-time concerning political party results in the districts of the electoral process.}\label{tbl:votes}
\begin{tabular}{l l p{8cm}}
\textbf{Variable}                & \textbf{Type}    & \textbf{Description} \\
\hline
\textbf{time}            & \textit{timestamp}  & Date and time of the data acquistion \\
\textbf{District}            & \textit{string}  & Short name of the location (district or nation-wide) \\
\textbf{Party}            & \textit{string}  & Political Party \\
\textbf{Mandates}            & \textit{numeric}  & MP's elected at the moment for the party in a given district\\
\textbf{Percentage}            & \textit{numeric}  & Percentage of votes in a party\\
\textbf{validVotesPercentage}            & \textit{numeric}  & Percentage of valid votes in a party\\
\textbf{Votes}            & \textit{numeric}  & Percentage of party votes\\
\hline
\end{tabular}
\end{center}
\end{table}

\subsubsection{Offline Acquisition}

Although not included in the Real-time Election Results data set, data on the finest granularity available (parishes) of the results is also acquired. In order to accomplish such task, a methodology similar to the online acquisition process was carried out, with the following difference: we query the respective site for the JSON files of each parish. 

There is no information on the website of the available local identifiers in bulk. As such, instead of an exhaustive search, we resorted to the information divulged by the Portuguese Government's Open Data initiative. This repository contains information concerning Portuguese parishes (see \url{https://dados.gov.pt/pt/datasets/freguesias-de-portugal/}), including their identification. Luckily, this is the same information used to identify locations in the election results website. As such, we were able to reduce the search from 490000 queries (from 010000, the lowest known identifier, to 500000, the national level identifier), down to 3092, the exact number of parishes. This process was carried out in the 3$^rd$ of December 2019.

The attributes of each file (\textit{i)} the overall election information and \textit{ii)} the voting information) are the same as in the online acquisition of data, with the following differences (strike-through for removed attributes):

\begin{itemize}
    \item To the overall election information:
    \begin{itemize}
        \item \textbf{Council} \textit{(string)}: The council of the parish;
        \item \textbf{District} \textit{(string)}: The district of the parish;
    \end{itemize}
    \item To the voting information:
    \begin{itemize}
        \item \textbf{Council} \textit{(string)}: The council of the parish;
        \item \textbf{District} \textit{(string)}: The district of the parish;
        \item \sout{\textbf{time} \textit{(timestamp)}}: Removed;
        \item \sout{\textbf{Mantes} \textit{(numeric)}}: Removed;
    \end{itemize}
\end{itemize}

Parish-level data provides a deeper understanding of the overall results and allows the introduction of several dimensions of the phenomena surrounding and influencing such results. Additionally, although not carried out at this point, this data also allows the addition of results from previous parliamentary elections in Portugal, under the assumption that the sequence of incoming results from parishes is the same.

We should clarify, again, that this data is made available jointly with the data set. Still, it was not used (does not belong to the group of raw data, which is only district-level) in the development of the final data set.


\section{Data Curation and Description}\label{sec:cur_and_description}

In this section, we describe the curation process applied to the raw data described in Section~\ref{sec:dataacquisition} and the final data set.

\subsection{Curation}\label{subsec:curation}

The process of data curation consists of the analysis of the information w.r.t consistency, data input/format errors, and related issues that may affect the future use of the data. In this case, we include the process of feature engineering under this concept.

An analysis of the overall and party voting data shows common issues concerning format errors and missing values. Concerning format errors, timestamps are formatted as such (previously as strings), and numerical variables were also cast accordingly. Concerning missing values, we observed that in the early stages of electoral results becoming available, individual districts did not have any information yet. As such, their attributes were marked as not available. Such cases are excluded from the analysis.

Concerning feature engineering, it was decided to add three features that could potentially help future users of the data set. First, the time elapsed (\textbf{TimeElapsed}) since the first data acquisition in minutes (\textit{numeric}); second, the application of the Hondt method (\textbf{Hondt}) to the results at each data acquisition step, at a district-level, resulting in their respective number of parliament members (\textit{numeric}); and third, the target variable (\textbf{FinalMandates}) with the final number of elected members of parliament for each district (\textit{numeric}).

We should note that some of the available attributes are excluded given their lack of relevance (at least apparently) to the possible predictive modelling tasks. The attributes removed are the following: \textbf{pre.availableMandates}, \textbf{pre.totalMandates}, \textbf{territoryFullName}, \textbf{territoryKey}. It should be reminded that at this point, we are focusing on the objective of predictive modelling. As such, repeated identifiers and redundant information (as these examples) should be removed.

Finally, the data concerning the overall election and the party voting information are joined w.r.t the timestamp of data acquisition and the respective district.

\subsection{Description}\label{subsec:description}

For the sake of consistency, the type of data in each attribute of the final data set on the election results for the Portuguese Parliament in 2019 is presented in Table~\ref{tbl:finalats}

\begin{table}[h]
\begin{center}
\scriptsize
\caption{Attributes of the Real-time Election Results - 2019 Portuguese Parliament Election.}\label{tbl:finalats}
\begin{tabular}{l l p{8cm}}
\textbf{Variable}                & \textbf{Type}    & \textbf{Description} \\
\hline
\textbf{TimeElapsed}            & \textit{Numeric}  & Time (minutes) passed since the first data acquisition \\
\textbf{time}            & \textit{timestamp}  & Date and time of the data acquisition \\
\textbf{territoryName}            & \textit{string}  & Short name of the location (district or nation-wide) \\
\textbf{totalMandates}            & \textit{numeric}  & MP's elected at the moment\\
\textbf{availableMandates}            & \textit{numeric}  & MP's left to elect at the moment\\
\textbf{numParishes}            & \textit{numeric}  & Total number of parishes in this location\\
\textbf{numParishesApproved}            & \textit{numeric}  & Number of parishes approved in this location\\
\textbf{blankVotes}            & \textit{numeric}  & Number of blank votes\\
\textbf{blankVotesPercentage}            & \textit{numeric}  & Percentage of blank votes\\
\textbf{nullVotes}            & \textit{numeric}  & Number of null votes\\
\textbf{nullVotesPercentage}            & \textit{numeric}  & Percentage of null votes\\
\textbf{votersPercentage}            & \textit{numeric}  & Percentage of voters\\
\textbf{subscribedVoters}            & \textit{numeric}  & Number of subscribed voters in the location\\
\textbf{totalVoters}            & \textit{numeric}  & Percentage of blank votes\\
\textbf{pre.blankVotes}            & \textit{numeric}  & Number of blank votes (previous election)\\
\textbf{pre.blankVotesPercentage}            & \textit{numeric}  & Percentage of blank votes (previous election)\\
\textbf{pre.nullVotes}            & \textit{numeric}  & Number of null votes (previous election)\\
\textbf{pre.nullVotesPercentage}            & \textit{numeric}  & Percentage of null votes (previous election)\\
\textbf{pre.votersPercentage}            & \textit{numeric}  & Percentage of voters (previous election)\\
\textbf{pre.subscribedVoters}            & \textit{numeric}  & Number of subscribed voters in the location (previous election)\\
\textbf{pre.totalVoters}            & \textit{numeric}  & Percentage of blank votes (previous election)\\
\textbf{Party}            & \textit{string}  & Political Party \\
\textbf{Mandates}            & \textit{numeric}  & MP's elected at the moment for the party in a given district\\
\textbf{Percentage}            & \textit{numeric}  & Percentage of votes in a party\\
\textbf{validVotesPercentage}            & \textit{numeric}  & Percentage of valid votes in a party\\
\textbf{Votes}            & \textit{numeric}  & Percentage of party votes\\
\textbf{Hondt}            & \textit{numeric}  & Number of MP's according to the distribution of votes now \\
\textbf{FinalMandates}            & \textit{numeric}  & \textbf{Target}: final number of elected MP's in a district/national-level \\
\hline
\end{tabular}
\end{center}
\end{table}


\section{User Notes}\label{sec:usernotes}

Basic operations in the programming language \textbf{R} are made available to facilitate the use of the data set. This file is located in the UCI data set archive.


\section*{Acknowledgments}\label{sec:ack}

National Funds finance this work through the Portuguese funding agency, FCT - Fundação para a Ciência e a Tecnologia within project UID/EEA/50014/2019.



\end{document}